\documentclass[12pt]{article}
\usepackage{amsmath,amsfonts,amssymb}
\usepackage{hyperref}
\usepackage{graphicx}
\usepackage{color}
\usepackage{epsfig}
\usepackage{psfrag}
\usepackage{ulem}

\usepackage{color}

\newcommand \be{\begin{eqnarray}}
\newcommand \ee{\end{eqnarray}}

\makeatletter\@addtoreset{equation}{section}\makeatother

\DeclareMathOperator{\Tr}{Tr}
\DeclareMathOperator{\tr}{tr}

\DeclareMathOperator{\re}{Re}

\def\bC {\mathbb{C}}

\def\bH {\mathbb{H}}

\def\bR {\mathbb{R}}

\def\bZ {\mathbb{Z}}

\newcommand{\beq}{\begin{equation}}
\newcommand{\eeq}{\end{equation}}
\newcommand{\bal}{\begin{equation}\begin{aligned}}
\newcommand{\eal}{\end{aligned} \end{equation}}
\newcommand{\bea}{\begin{eqnarray}}
\newcommand{\eea}{\end{eqnarray}}

\newcommand{\eqn}[1]{(\ref{#1})}
\newcommand{\app}[1]{Appendix~\ref{app:#1}}
\newcommand{\sect}[1]{Section~\ref{sec:#1}}

\newcommand{\address}[1]{\vbox{\center\em#1}}
\renewcommand{\title}[1]{\vbox{\center\huge{#1}}\vspace{5mm}}

\newcommand{\cJ}{{\mathcal J}}

\newcommand{\cL}{{\mathcal L}}

\newcommand{\cN}{{\mathcal N}}

\newcommand{\su}{\mathfrak{s}\mathfrak{u}}
\renewcommand{\sl}{\mathfrak{s}\mathfrak{l}}

\begin{document}
\bibliographystyle{utphys}

\begin{titlepage}
\begin{center}
\phantom{ }

\vspace{15mm}

\title{From algebraic curve to minimal surface \\and back}
\vspace{5mm}

\renewcommand{\thefootnote}{$\alph{footnote}$}

Michael Cooke\footnote{\href{mailto:cookepm@tcd.ie}{\tt cookepm@tcd.ie}} and Nadav Drukker\footnote{\href{mailto:nadav.drukker@gmail.com}{\tt nadav.drukker@gmail.com}}
\vskip 5mm
\address{
Department of Mathematics, King's College London \\
The Strand, WC2R 2LS, London, UK}

\renewcommand{\thefootnote}{\arabic{footnote}}
\setcounter{footnote}{0}

\end{center}

\vskip5mm

\abstract{
\normalsize
\noindent
We derive the Lax operator for a very large family of classical minimal surface solutions in 
$AdS_3$ describing Wilson loops in $\cN=4$ SYM theory. These solutions, constructed 
by Ishizeki, Kruczenski and Ziama, are associated with a hyperellictic surface of odd genus. 
We verify that the algebraic curve derived from the Lax operator is indeed none-other than 
this hyperelliptic surface.
}

\end{titlepage}

\section{Introduction}
\label{sec:intro}

Integrability has led in recent years to great advances in the detailed understanding 
of the $AdS$/CFT correspondence \cite{mal_2} between free IIB superstring theory in 
$AdS_{5}\times S^{5}$ and $\mathcal{N}=4$ supersymmetric Yang-Mills (SYM) 
in the planar approximation. 
The most well-studied sector is the spectrum of single-trace operators in the field theory 
and of the dual closed strings. Other advancements are in calculating scattering 
amplitudes
, structure constants and
open string states.

Open strings may end on D-branes or extend till the boundary of space, in which case 
they are dual to Wilson loop operators in the field theory. Studying them is more involved 
than solving the usual spectral problem for closed (or open) strings; at the classical 
level, the boundary conditions are not periodic (or Dirichlet/Neumann for fixed 
directions), but vary arbitrarily along the boundary. Solutions are not classified only by 
initial data, but by a rich structure of boundary conditions. If the Wilson loop is 
made of light-like segments, the regularized action on the surface (though not its shape) 
can be computed by solving a set of TBA-like equations \cite{alday}. In this paper we 
focus on a different case, when the Wilson loop is space-like and piece together different 
integrability approaches that have been used to study them, verifying their consistency.

The algebraic curve approach \cite{kazakov} provides a framework 
to study classical closed strings by constructing the monodromy 
associated to a flat connection. This monodromy serves as a generating function 
of an infinite tower of conserved charges, including the energy. 
Solutions can be classified in terms of a spectral curve, which is naturally 
defined from the monodromy matrix.

The analog story for open strings dual to Wilson loops is far less well understood.  
There are no non-trivial cycles on the world-sheet, which makes it harder to 
construct a monodromy matrix. In fact it is often stated that it is impossible 
to construct one, when actually this can be done by either considering an open 
boundary-to-boundary monodromy, or by taking the monodromy including reflections 
from two boundary points. This latter approach was indeed implemented  in \cite{bajnok} 
(based on \cite{oz}) for open strings ending on D-branes in $AdS_5 \times S^5$. Still, the analog 
construction for open strings ending on the boundary has not been found, so 
there is no satisfactory monodromy matrix for the strings dual to Wilson loops.

Complimentary approaches to the study of such surfaces were proposed in \cite{janik} and 
\cite{dekel}. In the theory of integrable systems, the monodromy of an infinite-dimensional 
system plays an analogous role to that of the Lax operator in finite-dimensional systems. 
In particular one may extract an algebraic curve from the Lax operator, as the spectral 
curve is extracted from the monodromy. 
One may however also construct a Lax operator for infinite-dimensional systems when 
restricting oneself to a finite subsystem. This Lax operator is local, in contrast to a monodromy. 
As such, it is not unique and \emph{\`a priori} neither is the resulting algebraic curve. 
Furthermore, the prescription for generating the conserved quantities from this operator 
is typically unclear. The constructions of \cite{janik,dekel} give an algebraic curve 
based on a Lax operator for classical strings in $AdS_{3}$. In this paper we extend their 
construction from genus-one solutions to higher genus.

We implement this on a very large class of explicit solutions in Euclidean $AdS_{3}$ 
which were constructed in \cite{ishizeki} and \cite{kski} 
(see also \cite{krucz_14}, for a different approach to these solutions). 
These solutions are expressed in 
terms of theta functions associated with a hyperelliptic curve of arbitrary odd genus. Furthermore, they 
depend on a spectral parameter, such that each curve gives rise to a one-parameter 
family of classical solutions.%
\footnote{There is a further discrete choice of possible boundaries for each curve 
and spectral parameter, see \cite{ishizeki}, but we shall not discuss it.}
The purpose of our study is to verify that the algebraic curve associated with these 
surfaces is indeed the hyperelliptic curve they were constructed from and 
elucidate the role of the spectral parameter.

We adopt a modified version of the approach in \cite{janik} to construct the Lax operator 
for the full family of solutions in \cite{ishizeki,kski} and derive the algebraic curve from it. Specifically, 
we construct a Lax operator whose Lax connection is the Pohlmeyer reduced connection 
rather than the sigma-model connection. Furthermore, we use the spectral 
parameter of \cite{ishizeki}, rather than the sigma-model spectral parameter used in \cite{janik}.

The structure of the paper is as follows: We begin in \sect{coset} by introducing the 
coset manifold description of Euclidean $AdS_{3}$. As the solutions of \cite{ishizeki} 
are expressed in terms of Riemann theta functions, we devote  \sect{theta} to 
introducing these functions and some of their properties. In \sect{all_g}, we 
construct the Lax pair and derive from it the 
algebraic curve. Finally, we demonstrate the construction 
for genus-one in \sect{g_1}.

While most of the discussion is self-contained, we chose for brevity not to review 
all the details of the construction in \cite{ishizeki,kski} and refer the reader there 
for where we have not. Our notation is mostly the same, with some exceptions 
which should not cause too much trouble for the reader.

\section{Euclidean $AdS_{3}$ sigma model}
\label{sec:coset}

We briefly review here the construction of the Euclidean $AdS_{3}$ ($\bH_{3}$) 
sigma-model. 
The integrability of this subsector of the $AdS_5\times S^5$ sigma model is most manifest when 
considered as the coset manifold $SL(2;\mathbb{C})/SU(2)$. It is parametrized by 
the group elements
\beq
g=
\begin{pmatrix}
X_{0}+X_{3} & X_{1}-iX_{2}\\
X_{1}+iX_{2} & X_{0}-X_{3}
\end{pmatrix}\in SL(2;\mathbb{C})/SU(2)\,,
\eeq
where $X_{i}$ are the embedding coordinates of $\bH_{3}$ in $\mathbb{R}^{1,3}$. 
The $SU(2)$ factor relates to the hermiticity of $g$. The sigma-model action is 
written in terms of $\sl(2;\bC)/\su(2)$ Maurer-Cartan forms as
\beq
\label{sig-action}
S=\frac{1}{4\pi}\int d^{2}w\,\tr(\cJ\bar{\cJ})\,,
\eeq
where $w$ and $\bar{w}$ are complex 
world-sheet coordinates (with $\partial$ and $\bar{\partial}$ their respective derivatives) 
and $\cJ=g^{-1}\partial g$ and $\bar{\cJ}=g^{-1}\bar{\partial}g$.
The sigma-model equations of motion and Virasoro constraints 
can then be written in terms of these currents.

Instead of using these currents, we choose, following \cite{ishizeki}, to employ the 
Pohlmeyer reduction of this model \cite{pohlmeyer}. 
Due to the hermiticity of $g$, we may perform the decomposition
\beq\label{decomp}
g=hh^{\dagger}\,,\qquad h\in SL(2;\mathbb{C})\,.
\eeq
Two $\sl(2;\mathbb{C})$ currents may naturally be associated with $h$ via 
$j\equiv h^{-1}\partial h$ and $\bar{j}\equiv h^{-1}\bar{\partial}h$. As a consequence 
of their definition, the Virasoro constraints and the equations of motion, 
These currents satisfy
\begin{subequations}
\begin{align}
\label{j_flat}
&\Tr j=\Tr \bar j=0\,,\qquad\bar\partial j-\partial\bar{j}+\left[\bar j,j\right]=0\,,
\\
\label{virasoro}
&\det(\bar j+j^\dagger)=0\,,
\\
\label{eom}
&\partial(\bar j+j^\dagger)+\frac{1}{2}\left[j-\bar j^\dagger,\bar j+j^\dagger\right]=0\,.
\end{align}
\end{subequations}

A \emph{spectral parameter}, $x\in\mathbb{C}$ may be introduced, as in \cite{janik,dekel}, 
to define new currents
\beq\label{J}
J(x)=\frac{1}{1-x}\left(j+x \bar j^\dagger\right)\,,
\qquad
\bar{J}(x)=\frac{1}{1+x}\left(\bar j-xj^\dagger\right)\,,
\eeq
such that the equations for $j$ and $\bar{j}$, \eqn{j_flat} and \eqn{eom}, 
are equivalent to the flatness condition for $J$ and $\bar{J}$
\beq\label{flat_con}
\bar\partial J-\partial\bar{J}+\left[\bar J,J\right]=0\,.
\eeq
While it is clear that \eqn{j_flat} is satisfied by $J(x)$ and $\bar{J}(x)$ for all $x$, 
a key fact is that for $\re(x)=0$ they also 
solve the equations of motion, \eqn{eom}. Thus for different values of imaginary $x$ 
the currents $J(x)$ and $\bar J(x)$ represent a family of real physical solutions of 
the sigma-model and the spectral parameter is not merely a formal expansion parameter.

An explicit paramerization of the currents $j$ and $\bar{j}$ was 
given in \cite{ishizeki}
\beq\label{param}
j=
\begin{pmatrix}
-\frac{1}{2}\partial\alpha & e^{-\alpha}\\
e^{\alpha} & \frac{1}{2}\partial\alpha
\end{pmatrix},
\qquad
\bar{j}=
\begin{pmatrix}
\frac{1}{2}\bar{\partial}\alpha & e^{\alpha}\\
-e^{-\alpha} & -\frac{1}{2}\bar{\partial}\alpha
\end{pmatrix},
\eeq
where $\alpha(w,\bar{w})$ is a real-valued function of the world-sheet coordinates, 
satisfying the cosh-Gordon equation
\beq
\partial\bar{\partial}\alpha=2\cosh(2\alpha)\,.
\eeq
Substituting \eqn{param} into \eqn{J}, we find the currents
\beq\label{gen_cxn}
J=
\begin{pmatrix}
-\frac{1}{2}\partial\alpha & e^{-\alpha}\\
\frac{1+x}{1-x}e^{\alpha} & \frac{1}{2}\partial\alpha
\end{pmatrix}\,,
\qquad
\bar{J}=
\begin{pmatrix}
\frac{1}{2}\bar{\partial}\alpha & \frac{1-x}{1+x}e^{\alpha}\\
-e^{-\alpha} & -\frac{1}{2}\bar{\partial}\alpha
\end{pmatrix}\,.
\eeq
Indeed it was already noticed in \cite{ishizeki} that for $\lambda=\frac{1+x}{1-x}$ on the unit 
circle, this describes real solutions, which exactly corresponds to imaginary $x$.%
\footnote{The same form of spectral parameter was found previously. 
See {\em e.g.}, \cite{kazama}.} 
$x$ serves thus in a dual role: Shifts along the imaginary axis lead to different solutions, 
while expanding the flatness condition around any such point leads to the usual 
equations of motion.

The cosh-Gordon equation can be solved in terms of theta functions of hyperelliptic 
Riemann surfaces. Given such a solution and a choice of $x$ we have the currents and 
$J$ and $\bar J$ and can reconstruct the full solution by reversing the procedure above. 
We first identify the flatness condition in \eqn{flat_con} as the compatibility condition for the 
auxiliary linear problem
\beq\label{alp}
\partial\psi=\psi J\,,\qquad\bar{\partial}\psi=\psi\bar{J}\,.
\eeq
For imaginary $x$ this connection also satisfies the equations of motion \eqn{eom} so
we may then identify $h(x)$ 
in \eqn{decomp} 
with the matrix made of the two linearly independent solutions 
to the auxiliary problem. From $h(x)$, we can then construct the group element $g(x)$ 
\eqn{decomp}, giving the solution to the $\bH_{3}$ sigma-model. Thus, 
changing the spectral parameter enables us to construct from $J$ and $\bar{J}$ 
a solution given by 
the group element 
$g(x)$ 
\emph{at any point along the imaginary-axis}.

\section{Riemann theta functions}
\label{sec:theta}

The procedure outlined above 
gives
solutions to the $\bH_3$ 
sigma-model from solutions of the cosh-Gordon equation. The fact that this equation 
can be solved by theta functions associated with hyperelliptic Riemann surfaces of 
odd genus \cite{babich} allows to find a very large family of solutions to the sigma-model 
\cite{ishizeki}. We provide a quick review 
to theta functions, focusing on properties relevant for our purposes. A lot more 
can be found in the abundant literature on the subject, {\em e.g.}, 
\cite{mfordi,mfordii,mfordiii,igusa,farkas}. 

The surface is a hyperelliptic curve defined by a function $f(\lambda)$ of the form%
\footnote{We will also refer to the function itself, and not only its solutions, as the curve.}
\beq\label{alg_c}
f(\lambda)=\sqrt{\lambda}\prod_{i=1}^{2g}\sqrt{\lambda-\lambda_{i}}\,,
\eeq
where $\lambda_i\in\mathbb{C}$, together with 
$\lambda=0$ and $\lambda=\infty$, are the branch points of $f$. It is apparent 
that such surfaces are two-sheeted, corresponding to the two branches of the square 
root. Furthermore, to ensures real solutions to the sigma-model, 
we require the surface to be equipped with an \emph{anti-holomorphic involution}, 
$\tau:\lambda\mapsto-1/\bar{\lambda}$, and in particular the set 
of $2g+2$ branch points $\{\lambda_{i}\}$ is invariant under this involution. 

For any a Riemann surface $\Sigma$ there exists a canonical basis of 
homology cycles $\{a_{i},b_{i}\}$ with $i=1,...,g$, satisfying 
$a_{i}\circ a_{j}=b_{i}\circ b_{j}=0$, and, $a_{i}\circ b_{j}=\delta_{ij}$. A set of 
cohomology one-forms $\{\omega_{i}\}$ dual to $\{a_{i}\}$, {\em i.e.}, satisfying 
$\oint_{a_{i}}\omega_{j}=\delta_{ij}$, may be defined. With this one constructs the 
\emph{period matrix}
\beq
\label{period}
\Pi_{ij}=\oint_{b_{i}}\omega_{j}\,.
\eeq
This is a symmetric $g\times g$ matrix with positive definite imaginary part. 

Riemann theta functions are defined as
\beq
\label{def_theta}
\theta(\Pi;z)
\equiv
\sum_{n\in\mathbb{Z}^{g}}\exp
\left(2\pi i\left(\frac{1}{2}n^{\intercal}\Pi n+n^{\intercal}z\right)\right),
\eeq
where $z\in\mathbb{C}^{g}$ and $\Pi$ is a $g\times g$ symmetric matrix with positive 
definite imaginary part. We consider the Riemann theta function associated with 
$\Sigma$ by taking $\Pi$ to be the period matrix \eqn{period}. We keep throughout a 
fixed Riemann surface and therefore write $\theta(z)\equiv\theta(\Pi;z)$.

The $g\times2g$ matrix $(I,\Pi)$ (where $I$ is the $g\times g$ identity matrix) generates 
a lattice, denoted by $\mathcal{L}(\Sigma)=(I,\Pi)\times\bZ^{2g}$. The \emph{Jacobian variety} of the surface $\Sigma$ is the quotient 
$\mathfrak{J}(\Sigma)=\mathbb{C}^{g}/\mathcal{L}(\Sigma)$. 
One may then define the \emph{Abel-Jacobi mapping}
\beq\label{abel}
\phi:\Sigma\rightarrow\mathfrak{J}(\Sigma)\,,\qquad\phi(\lambda)=\int_{\lambda_{0}}^{\lambda}\omega\,,
\eeq
where $\lambda_{0}$ is the \emph{base point} of the mapping, which we choose to be at 
$\lambda_{0}=0$. We often represent the 
Abel-Jacobi mapping by the integral $\int_{0}^{\lambda}\equiv\phi(\lambda)$.

Strictly, we should distinguish between points $p\in\Sigma$ and 
their projection $\lambda_p$ onto the complex plane. We may represent the uplift 
to the Riemann surface 
as $p=(\lambda, f(\lambda))$, and denote the other uplift $p'=(\lambda, -f(\lambda))$ 
(at branch points they are, of course, degenerate). The endpoint of integration in 
\eqn{abel} is really not $\lambda$ itself, but one of its uplifts. We will often 
use this notation and label corresponding points on the two sheets as $\lambda$ 
and $\lambda'$.

It is useful to define theta functions with \emph{characteristics}. Given 
$n,m\in\mathbb{C}^{g}$, the theta function with these characteristics is given by
\beq\label{char}
\theta\left[\begin{smallmatrix}
n\\
m
\end{smallmatrix}\right](z)
=e^{\pi i\left(\frac{1}{4}n^{\intercal}\Pi n+n^{\intercal}z+\frac{1}{2}n^{\intercal}m\right)}
\theta\left(z+\frac{1}{2}m+\frac{1}{2}\Pi n\right).
\eeq
We will concern ourselves only with integer characteristics $n,m\in\mathbb{Z}^{g}$. 
From \eqn{char} and (for the second equality) the definition \eqn{def_theta} we find
\beq\label{quasi}
\theta\left[\begin{smallmatrix}
2n\\
2m
\end{smallmatrix}\right](z)
=
e^{\pi i\left(n^{\intercal}\Pi n+2n^{\intercal}z+2n^{\intercal}m\right)}
\theta(z+m+\Pi n)
=\theta(z)\,.
\eeq
The theta functions are therefore quasi-periodic 
so they are defined, up to monodromies, on the Jacobian. 
An integer characteristic is said to be \emph{odd} or 
\emph{even} if the scalar product, $n^{\intercal}m$, is odd or even. 
We note that theta functions with odd characteristics are odd functions of 
$z\in\mathbb{C}^{g}$ and theta functions with even characteristics are even. 

Let us consider now the Abel-Jacobi map more explicitly. The differentials in 
\eqn{period} are given, up to a normalization, by \cite{farkas}
\beq
\label{explicit-omega}
\omega_{i}(\lambda)=\frac{\lambda^{i-1}d\lambda}{f(\lambda)}\,,
\eeq
with $f(\lambda)$ in \eqn{alg_c}. Since $f(\lambda')=-f(\lambda)$, then also 
$\omega(\lambda')=-\omega(\lambda)$ and consequently, 
$\int_{0}^{\lambda'}=-\int_{0}^{\lambda}$.

Despite the fact that the Abel-Jacobi mapping is not single-valued (it depends on the 
contour of integration), the composite map 
$\theta\circ\phi:\Sigma\rightarrow\mathbb{C}$ has a well-defined set of 
zeros. In particular, it has $g$ zeros and no poles \cite{farkas}. The same 
is clearly true with characteristics. 
There are $2g+2$ branch points of the algebraic curve defining the Riemann surface. 
To each of these branch points we may naturally associate a characteristic. 
The integral $\int_{0}^{\lambda_{i}}$ is half of the period integral between $0$ to 
$\lambda_{i}$ and back to 0 on the second sheet. Therefore it can be expressed 
as $\int_{0}^{\lambda_{i}}=\frac{1}{2}mI+\frac{1}{2}\Pi n$, for some 
$m,n\in\mathbb{Z}$. Now, applying \eqn{char}
\beq
\theta({\textstyle\int_{0}^{\lambda_{i}}})
=e^{\pi i\left(-\frac{1}{4}n^{\intercal}\Pi n-\frac{1}{2}n^{\intercal}m\right)}\theta\left[
\begin{smallmatrix}
n\\
m
\end{smallmatrix}
\right](0)\,,
\eeq
If $[\begin{smallmatrix}{n}\\m\end{smallmatrix}]$ 
is an odd characteristic, then $\theta[\begin{smallmatrix}{n}\\m\end{smallmatrix}](z)$ is an odd 
function and thus $\theta(\int_{0}^{\lambda_{i}})$ vanishes. 
We refer to branch points for which the corresponding characteristics are odd, as 
\emph{odd branch points} (and correspondingly for even characteristics). 
We denote by 
$\lambda^{-}_{i}$ the odd 
branch points and by 
$\lambda^{+}_{i}$ the even 
branch points. There are precisely $g$ odd branch points 
of a hyperelliptic surface, corresponding to the $g$ zeros of $\theta\circ\phi$ \cite{farkas}.
We fixed our curve to have branch points at 0 and at $\infty$ and a special role will be 
played by the theta functions with the appropriate characteristics. For $0$, this is the 
original theta function. The other one is defined as 
$\hat\theta(z)=\theta[\begin{smallmatrix}k\\l\end{smallmatrix}](z)$, where 
$2\int_{0}^{\infty}=l+\Pi k$. 

$2\int_{0}^{\lambda_{i}}=Im+\Pi n$ with $m,n\in\mathbb{Z}$ is a full period. Combining this with the fact that $\hat{\theta}(0)=0$ implies 
$\hat{\theta}(2\int_{0}^{\lambda})=0$. 
In fact for any integer $p$, a theta functions of the form 
$\theta[\begin{smallmatrix}{n}\\m\end{smallmatrix}](p\int_{0}^{\lambda})$, 
has $p^{2}g$ zeros (including multiplicities) \cite{farkas}. 
As we show in \app{zeros}, in the case of $\hat{\theta}(2\int_{0}^{\lambda})$ these 
are indeed just the $2g+2$ branch points, of which $g-1$ (the finite odd ones) 
are triple zeros and the other $g+3$ branch points are simple zeros on the Riemann surface.

Let us mention one more necessary tool: \emph{Directional derivatives} 
with respect to a point $\lambda$ on the Riemann surface are defined as
\beq
D_{\lambda}\theta(z)\equiv\omega(\lambda)\cdot\frac{\partial}{\partial z}\theta(z)\,.
\eeq

\section{The Lax operator and algebraic curve}
\label{sec:all_g}

Recall that a Lax operator is part of a Lax pair $(L,M)$, where the evolution of the 
system is described by
\beq\label{1d_lax}
\frac{dL}{dt}=[L,M],
\eeq
with $t$ some `time' parameter. For infinite-dimensional systems, such as the 
sigma-model, we may treat the two world-sheet coordinates as independent `time' 
parameters and study
\beq\label{lax_eqn}
dL=[L,M]\,,
\eeq
where $d$ is the exterior derivative with respect to the world-sheet coordinates of the open string, 
$d=dw\,\partial+d\bar{w}\,\bar{\partial}$.
An algebraic curve is naturally associated to this Lax operator via the characteristic polynomial
\beq
\det(L-yI)=0\,.
\eeq
For the $\mathbb{H}_{3}$ sector this simplifies to
\beq
	y^{2}=-\det L\,.
\eeq
The Lax equations ensure that this curve is independent the world-sheet position.

We now proceed to employ the method proposed in \cite{janik} (a variation of a 
theorem in \cite{babelon}) to find the Lax operator and from it the algebraic curve 
for the minimal surface solutions discussed above. 
It is natural that the result should be the algebraic curve \eqn{alg_c} associated 
to the theta function, but it is rather opaque how this 
would come about. The prescription in \cite{janik} is to take the Lax operator as
\beq\label{lax_op}
L(w,\bar{w};x)=\hat{\Psi}(w,\bar{w};x)^{-1}\cdot A(x)\cdot\hat{\Psi}(w,\bar{w};x)\,,
\eeq
where $\hat{\Psi}(w,\bar{w};x)$ is the matrix whose rows are the linearly independent 
solutions of an auxiliary linear problem of the form, 
\beq
\partial\psi=\psi \mathcal{J}(x)\,,\qquad\bar{\partial}\psi=\psi\bar{\mathcal{J}}(x)\,.
\eeq
and the matrix $A(x)$ is determined by the requirement that $L(w,\bar{w};x)$ have 
polynomial dependence on the spectral parameter. Here
\beq\label{gen_mc}
\mathcal{J}(x)=\frac{\mathcal{J}}{1-x}\,,\qquad\bar{\mathcal{J}}(x)=\frac{\bar{\mathcal{J}}}{1+x}\,,
\eeq
are defined 
in terms of the Maurer-Cartan forms $\mathcal{J}$ and $\bar{\mathcal{J}}$
in \eqn{sig-action}.

We wish to modify this prescription for our purposes. Firstly, we will construct a 
Lax operator with respect to the Pohlmeyer reduced connection $j$ and $\bar j$, 
{\em i.e.}, for $M=J(\lambda)dw+\bar{J}(\lambda)d\bar{w}$, 
rather than $M=\cJ(x)dw+\bar{\cJ}(x)d\bar{w}$. 
It is then natural to replace $x$ in the above argument by $\lambda=\frac{1+x}{1-x}$. 
While an algebraic curve in 
terms of $x$ is clearly related by a birational transformation to one in terms of 
$\lambda$ (and therefore structurally equivalent), the condition for polynomality 
of the Lax matrix in $\lambda$ or in $x$ are inequivalent. The sigma-model 
connection and the Pohlmeyer reduced connection are related by a gauge transformation 
\cite{kazama}, and we thus expect the resulting spectral curve to be the same. 
We demonstrate the equivalence of solving the respective Lax equations in \app{lax}. 

We shall proceed to construct the Lax operator and the corresponding algebraic curve. 
Beforehand, however, we need a further result from \cite{ishizeki}; 
the solutions to the auxiliary linear problem. Recall that $h(w,\bar{w};\lambda)$ is made of the two independent solutions and is given by \cite{ishizeki}
\beq\label{h}
\begin{aligned}
&h(w,\bar{w};\lambda)=\left(\frac{\theta^{2}(\int_{0}^{\lambda})}
{2\hat{\theta}(2\int_{0}^{\lambda})D_{\infty}\hat{\theta}(0)}\right)^{1/2}\\
&\times\begin{pmatrix}
\sqrt{-\lambda}\,\frac{\hat{\theta}(z+\int_{0}^{\lambda})}{\hat{\theta}(z)}
\left(\frac{\hat{\theta}(z)}{\theta(z)}\right)^{1/2}
e^{\mu(\lambda)w+\nu(\lambda)\bar{w}}&
\frac{\theta(z+\int_{0}^{\lambda})}{\theta(z)}
\left(\frac{\theta(z)}{\hat{\theta}(z)}\right)^{1/2}
e^{\mu(\lambda)w+\nu(\lambda)\bar{w}}\\
-\sqrt{-\lambda}\,\frac{\hat{\theta}(z-\int_{0}^{\lambda})}{\hat{\theta}(z)}
\left(\frac{\hat{\theta}(z)}{\theta(z)}\right)^{1/2}
e^{-\mu(\lambda)w-\nu(\lambda)\bar{w}}&
\frac{\theta(z-\int_{0}^{\lambda})}{\theta(z)}
\left(\frac{\theta(z)}{\hat{\theta}(z)}\right)^{1/2}
e^{-\mu(\lambda)w-\nu(\lambda)\bar{w}}
\end{pmatrix}\,,
\end{aligned}
\eeq
where $z=z(w,\bar{w})$ is a function of the world-sheet coordinates. The precise form 
of $z(w,\bar w)$, of $\mu(\lambda)$ and of $\nu(\lambda)$ 
will not be important for our discussion. As we expect the curve to be 
hyperelliptic, we make the ansatz \cite{babelon}
\beq
A(\lambda)=g(\lambda)
\begin{pmatrix}
1 & 0\\
0 & -1
\end{pmatrix}.
\eeq
The resulting Lax operator is given by
\beq\label{lax}
L(w,\bar{w};\lambda)\equiv
\begin{pmatrix}
L_{11} & L_{12}\\
L_{21} & L_{22}
\end{pmatrix}\,,
\eeq
where
\begin{subequations}
\label{L_all}
\begin{align}
\label{L11}
L_{11}&=-L_{22}
=\frac{\sqrt{-\lambda}\,\theta^{2}(\int_{0}^{\lambda})g(\lambda)}
{2\hat{\theta}(2\int_{0}^{\lambda})D_{\infty}\hat{\theta}(0)}
\left(\frac{\hat{\theta}(z+\int_{0}^{\lambda})\theta(z-\int_{0}^{\lambda})
-\theta(z+\int_{0}^{\lambda})\hat{\theta}(z-\int_{0}^{\lambda})}{\theta(z)\hat{\theta}(z)}\right),
\\
\label{L12}
L_{12}&=\frac{\theta^{2}(\int_{0}^{\lambda})g(\lambda)}
{\hat{\theta}(2\int_{0}^{\lambda})D_{\infty}\hat{\theta}(0)}
\frac{\theta(z+\int_{0}^{\lambda})\theta(z-\int_{0}^{\lambda})}{\theta(z)\hat{\theta}(z)}\,,
\\
\label{L21}
L_{21}&=-\frac{\lambda\theta^{2}(\int_{0}^{\lambda})g(\lambda)}
{\hat{\theta}(2\int_{0}^{\lambda})D_{\infty}\hat{\theta}(0)}
\frac{\hat{\theta}(z+\int_{0}^{\lambda})\hat{\theta}(z-\int_{0}^{\lambda})}{\theta(z)\hat{\theta}(z)}\,.
\end{align}
\end{subequations}
We now emphasize a point about this construction. It is apparent that $g(\lambda)$ 
must be the algebraic curve as, by the definition of $L$, $\det L=-g(\lambda)^{2}$. 
This observation advocates the converse procedure of that in \cite{janik}, {\em i.e.}, 
we make an ansatz for $g(\lambda)$, 
and we need to verify the polynomality of the resulting Lax matrix. In order 
to do that we need algebraic relations between $f(\lambda)$ 
in \eqn{alg_c}
and the theta functions.

\subsection{A meromorphic function}
\label{sec:mer}

Consider the function
on the Riemann surface $\Sigma$, given by
\beq\label{gfrak}
G(p)=\frac{\hat{\theta}(2\int_{0}^{\lambda_{p}})}{\theta^{4}(\int_{0}^{\lambda_{p}})}\,.
\eeq
Recall that $p=(f(\lambda_{p}),\lambda_{p})$, where 
$\lambda_{p}$ is the projection of $p$ onto the complex plane. First note that 
it is single valued on $\Sigma$
and in 
particular, it is independent of the contour 
of integration. To see that, consider the monodromy around a closed cycle 
$\int_{0}^{\lambda_{p}}\rightarrow\int_{0}^{\lambda_{p}}+m+\Pi n$, for 
$m,n\in\mathbb{Z}$. Applying \eqn{quasi} and \eqn{char} we find
\beq
G(p)\to\frac{\hat{\theta}(2\int_{0}^{\lambda_{p}}+2m+2\Pi n)}
{\theta^{4}(\int_{0}^{\lambda_{p}}+m+\Pi n)}
=\frac{\hat{\theta}(2\int_{0}^{\lambda_{p}})}{\theta^{4}(\int_{0}^{\lambda_{p}})}\,,
\eeq
Since both $\theta$ and $\hat\theta$ are analytic, $G(p)$ is meromorphic on $\Sigma$.

If instead we consider $G$ as a function on the complex $\lambda$-plane, it will have 
the same branch-cuts as $f(\lambda)$, and then $G^2(\lambda)$ is meromorphic on the 
complex $\lambda$-plane.

One may be tempted to guess that $G(\lambda)=f(\lambda)$, but this is not the case. 
The zeros of the numerator and denominator have already been discussed in \sect{theta} 
and \app{zeros}. Combining these results, we see that $G^2$ has%
\footnote{Since all the zero and poles are branch points of $G(\lambda)$, the same 
statements actually hold for $G$ as a function on $\Sigma$.}
\begin{itemize}
\item simple zeros at at the $g+2$ even branch points, $\{\lambda^{+}_{i}\}$,
\item simple poles at the $g-1$ odd finite branch points, $\{\lambda^{-}_{i}\}$,
\item a third order pole at $\infty$,
\item no other zeros or poles.
\end{itemize}
We may therefore identify $G$ as
\beq
G(\lambda)
=G_0\frac{\prod_{i}^{g+2}\sqrt{\lambda-\lambda^{+}_{i}}}
{\prod_{i}^{g-1}\sqrt{\lambda-\lambda^{-}_{i}}}\,.
\eeq

We can determine the constant, though it is not really crucial for our purposes, 
from the asymptotics of the theta functions
\beq
\lim_{\lambda\rightarrow\infty}\lambda^{-3/2}G(\lambda)=\frac{1}{4}
\left(\frac{1}{D_{\infty}\hat{\theta}(0)}\right)^{3}\,.
\eeq
Recall that \cite{ishizeki}
\beq\label{spec_p}
\sqrt{-\lambda}=2\frac{D_{\infty}\hat{\theta}(0)\,\hat{\theta}(\int_{0}^{\lambda})}
{\theta(0)\theta(\int_{0}^{\lambda})}\,,
\eeq
So we find that
\beq
G_0=\frac{D_{0}\hat{\theta}(0)}{\left(\theta(0)D_{\infty}\hat{\theta}(0)\right)^{2}}\,.
\eeq

Splitting the product \eqn{alg_c} into terms with the odd finite branch points $\{\lambda^{-}_{i}\}$ 
and with the even branch points $\{\lambda^{+}_{i}\}$ we find
\beq\label{alg_c_exp}
 g(\lambda)
 =G(\lambda)\prod_{i}^{g-1}\left(\lambda-\lambda^{-}_{i}\right)
=\sqrt{\lambda}\prod_{i}^{g-1}\sqrt{\lambda-\lambda^{-}_{i}}\prod_{j}^{g+1}\sqrt{\lambda-\lambda^{+}_{j}}
= f(\lambda)\,.
\eeq
So while $G(\lambda)$ is not equal to $f(\lambda)$, their ratio is a polynomial.

We now use this representation of $f(\lambda)$ to prove that the entries of the Lax 
matrix \eqn{L_all} are indeed polynomials in $\lambda$ and find the order of these polynomials.
\subsection{$L_{11}$}
Using $\int_0^{\lambda'}=-\int_0^\lambda$ where the integrals are along image paths on the 
two sheets, we can rewrite \eqn{L11} as
\beq
\label{L11_final}
L_{11}=\frac{\theta^{2}(0)D_{\infty}\hat{\theta}(0)}{2D_{0}\hat{\theta}(0)}
\frac{\hat{\theta}(z+\int_{0}^{\lambda})\theta(z-\int_{0}^{\lambda})
-\hat{\theta}(z+\int_{0}^{\lambda'})\theta(z-\int_{0}^{\lambda'})}
{\theta(z)\hat{\theta}(z)}
\sqrt{-\lambda}\,
\frac{\prod_{i}^{g-1}(\lambda-\lambda^{-}_{i})}{\theta^{2}(\int_{0}^{\lambda})}\,,
\eeq
\begin{itemize}
\item{\bf No branch cuts:} 
Clearly the $\sqrt{-\lambda}$ term introduces a branch cut which changes sign between 
the two sheets $\lambda\rightarrow\lambda'$, but this is negated by the branch cut of the 
second fraction (see \app{tay}). Therefore overall $L_{11}$ is single valued.
\item{\bf Analyticity:} 
As discussed in \sect{theta}, the last denominator has zeros at the odd finite branch 
points and at infinity. These are simple zeros which are canceled by 
corresponding zeros of the polynomial in the numerator.
\item{\bf Order:}
Near the origin the two terms with branch cuts in \eqn{L11_final} mean that $L_{11}(0)=0$, 
so $L_{11}$ has no constant piece. At large $\lambda$, it is shown in \app{tay} that the 
second fraction scales together like $\lambda^{-1/2}$, as does $\theta(\int_0^\lambda)$. 
We find that $L_{11}=O(\lambda^g)$ and is therefore a degree $g$ polynomial in $\lambda$.
\end{itemize}

\subsection{$L_{12}$}
\beq
L_{12}=\frac{\theta^{2}(0)D_{\infty}\hat{\theta}(0)}{D_{0}\hat{\theta}(0)}
\frac{\theta(z+\int_{0}^{\lambda})\theta(z-\int_{0}^{\lambda})}{\theta(z)\hat{\theta}(z)}
\frac{\prod_{i}^{g-1}(\lambda-\lambda^{-}_{i})}{\theta^{2}(\int_{0}^{\lambda})}\,,
\eeq
\begin{itemize}
\item{\bf No branch cuts:} 
$L_{12}$ is invariant under $\lambda\rightarrow\lambda'$.
\item{\bf Analyticity:} 
The only possible zeros of the denominator are at $\lambda^{-}_{i}$ and are cancelled 
by those in the numerator.
\item{\bf Order:}
The $z$-dependant fraction has a finite limit at large $\lambda$. The last one scales 
like $\lambda^g$, so $L_{12}$ is a polynomial of degree $g$.
\end{itemize}

\subsection{$L_{21}$}
\beq
L_{21}=-\frac{\theta^{2}(0)D_{\infty}\hat{\theta}(0)}{D_{0}\hat{\theta}(0)}
\frac{\hat{\theta}(z+\int_{0}^{\lambda})\hat{\theta}(z-\int_{0}^{\lambda})}{\theta(z)\hat{\theta}(z)}
\frac{\lambda\prod_{i}^{g-1}(\lambda-\lambda^{-}_{i})}{\theta^{2}(\int_{0}^{\lambda})}\,,
\eeq
\begin{itemize}
\item
The same arguments as in the case of $L_{12}$ show that $L_{21}$ is a polynomial of degree 
$g+1$ with no constant term.
\end{itemize}

We conclude that with the representation \eqn{alg_c_exp} of $f(\lambda)$, the Lax operator 
$L(w,\bar w,\lambda)$ given by \eqn{L_all} is indeed the Lax operator describing the family 
of solutions and its determinant gives 
the 
corresponding 
algebraic curve.
This implies that the procedure advocated in \cite{janik} indeed applies to the 
open string solutions of \cite{ishizeki} 
with the spectral parameter
$\lambda=(1+x)/(1-x)$ and as expected, the 
resulting curve is none other than the hyperelliptic curve defining the Riemann theta functions.

\section{The genus-one case}
\label{sec:g_1}

We have constructed the Lax operator for a solution based on an arbitrary genus curve, 
shown that its entries are polynomial in $\lambda$ and proven that the resulting 
algebraic curve is indeed $f(\lambda)$ \eqn{alg_c}. We have not presented, though, 
explicit expressions for the coefficients of 
the polynomials in 
$\lambda$ in the 
different matrix elements of $L$. To do that would require to disentangle expressions 
like $\theta(z+\int_0^\lambda)\theta(z-\int_0^\lambda)$. 
In fact the addition theorem 
(see {\em e.g.}, \cite{igusa}) may be employed to do just that. 
This allows one to write such products as sums over 
products of theta functions of $z$ and theta functions of $\lambda$, thus splitting the 
spectral parameter and world-sheet dependence in \eqn{L_all}. It does so at the 
expense of introducing a sum over all possible integer (mod $2$) characteristics. 
We have not found the resulting expressions for arbitrary genus illuminating, but 
in the case of genus-one, which we present here, they are very explicit and 
rather simple

For genus-one the Riemann theta functions reduce to elliptic theta 
functions, see \cite{kski}
\beq
\theta(z)=\vartheta_{3}(\pi z;q)\,,\qquad\hat{\theta}(z)
=-\vartheta_{1}(\pi z;q)\,,\qquad\text{with}\quad q=e^{i\pi\Pi}\,.
\eeq
The period matrix is given by $\Pi=i\mathbb{K}(k^{\prime})/\mathbb{K}(k)$, where $\mathbb{K}$ is the complete elliptic integral of the first kind, and $k$ and $k^{\prime}$ are the elliptic modulus and complementary elliptic modulus, respectively.

Employing the result for $f(\lambda)$ from \sect{all_g}, we have
\beq
f(\lambda)=\sqrt{\lambda}\sqrt{\lambda-\lambda_1}\sqrt{\lambda-\lambda_2}
=-\vartheta_{3}^{2}(0)D_{\infty}\vartheta_{1}(0)
\frac{\vartheta_{1}(2\int_{0}^{\lambda})}{\vartheta_{3}^{4}(\int_{0}^{\lambda})}\,,
\eeq
where $\lambda_2=-1/\bar\lambda_1$. 
We have absorbed a factor of $\pi$ in the definition of the elliptic functions 
and used that for genus-one $\omega(0)=-\omega(\infty)$. Note that the cross-ratio 
of $\{0,\infty,\lambda_1,-1/\bar\lambda_1\}$ is always real, so they are all along a line, 
which without loss of generality we take to be the real line. Furthermore the only odd 
branch point is at infinity. We also note that the elliptic modulus and complementary 
modulus may be expressed in terms of the branch points via
\beq
k=\frac{a}{\sqrt{1+a^{2}}}\,,
\qquad
k^{\prime}=\frac{1}{\sqrt{1+a^{2}}}\,.
\eeq

The expression for the Lax operator, \eqn{L_all}, becomes
\bal
L_{11}&=-\frac{\sqrt{-\lambda}}{2}\vartheta_{3}^{2}(0)
\frac{\vartheta_{1}(z+\int_{0}^{\lambda})\vartheta_{3}(z-\int_{0}^{\lambda})
-\vartheta_{3}(z+\int_{0}^{\lambda})\vartheta_{1}(z-\int_{0}^{\lambda})}
{\vartheta_{3}(z)\vartheta_{1}(z)\vartheta_{3}^{2}(\int_{0}^{\lambda})}\,,
\\
L_{12}&=\vartheta_{3}^{2}(0)
\frac{\vartheta_{3}(z+\int_{0}^{\lambda})\vartheta_{3}(z-\int_{0}^{\lambda})}
{\vartheta_{3}(z)\vartheta_{1}(z)\vartheta_{3}^{2}(\int_{0}^{\lambda})}\,,\\
L_{21}&=-\lambda\vartheta_{3}^{2}(0)
\frac{\vartheta_{1}(z+\int_{0}^{\lambda})\vartheta_{1}(z-\int_{0}^{\lambda})}
{\vartheta_{3}(z)\vartheta_{1}(z)\vartheta_{3}^{2}(\int_{0}^{\lambda})}\,.
\eal
We would like to evaluate the coefficients of the polynomials, and to explicitly recover the spectral curve. For genus-one, the expression for the spectral parameter \eqn{spec_p} simplifies to
\beq
\lambda=\frac{\vartheta_{1}^{2}(\int_{0}^{\lambda})}{\vartheta_{3}^{2}(\int_{0}^{\lambda})}\,.
\eeq

The addition theorem at genus-one is particularly simple and may be employed to split the 
world-sheet and spectral parameter dependence
\beq
\begin{gathered}
{\textstyle\vartheta_{1}(z+\int_{0}^{\lambda})\vartheta_{3}(z-\int_{0}^{\lambda})
-\vartheta_{3}(z+\int_{0}^{\lambda})\vartheta_{1}(z-\int_{0}^{\lambda})}
=2\frac{\vartheta_{2}(z)\vartheta_{4}(z)\vartheta_{1}(\int_{0}^{\lambda})\vartheta_{3}(\int_{0}^{\lambda})}
{\vartheta_{2}(0)\vartheta_{4}(0)}\,,
\\
{\textstyle\vartheta_{3}(z+\int_{0}^{\lambda})\vartheta_{3}(z-\int_{0}^{\lambda})}
=\frac{\vartheta_{1}^{2}(z)\vartheta_{1}^{2}(\int_{0}^{\lambda})
+\vartheta_{3}^{2}(z)\vartheta_{3}^{2}(\int_{0}^{\lambda})}
{\vartheta_{3}^{2}(0)}\,,
\\
{\textstyle\vartheta_{1}(z+\int_{0}^{\lambda})\vartheta_{1}(z-\int_{0}^{\lambda})}
=\frac{\vartheta_{1}^{2}(z)\vartheta_{3}^{2}(\int_{0}^{\lambda})
-\vartheta_{3}^{2}(z)\vartheta_{1}^{2}(\int_{0}^{\lambda})}{\vartheta_{3}^{2}(0)}\,.
\end{gathered}
\eeq

This gives
\bal
\label{g_1_fin}
L_{11}&=-\sqrt{-\lambda}\frac{\vartheta_{3}^{2}(0)}{\vartheta_{2}(0)\vartheta_{4}(0)}\frac{\vartheta_{2}(z)\vartheta_{4}(z)\vartheta_{1}(\int_{0}^{\lambda})}{\vartheta_{3}(z)\vartheta_{1}(z)\vartheta_{3}(\int_{0}^{\lambda})}=\mp i\frac{\lambda}{\sqrt{kk^{\prime}}}\frac{\vartheta_{2}(z)\vartheta_{4}(z)}{\vartheta_{3}(z)\vartheta_{1}(z)}\,,\\
L_{12}&=\frac{\vartheta_{1}^{2}(z)\vartheta_{1}^{2}(\int_{0}^{\lambda})+\vartheta_{3}^{2}(z)\vartheta_{3}^{2}(\int_{0}^{\lambda})}{\vartheta_{3}(z)\vartheta_{1}(z)\vartheta_{3}^{2}(\int_{0}^{\lambda})}=\lambda\frac{\vartheta_{1}(z)}{\vartheta_{3}(z)}+\frac{\vartheta_{3}(z)}{\vartheta_{1}(z)}\,,\\
L_{21}&=-\lambda\frac{\vartheta_{1}^{2}(z)\vartheta_{3}^{2}(\int_{0}^{\lambda})-\vartheta_{3}^{2}(z)\vartheta_{1}^{2}(\int_{0}^{\lambda})}{\vartheta_{3}(z)\vartheta_{1}(z)\vartheta_{3}^{2}(\int_{0}^{\lambda})}=\lambda^{2}\frac{\vartheta_{3}(z)}{\vartheta_{1}(z)}-\lambda\frac{\vartheta_{1}(z)}{\vartheta_{3}(z)}\,,
\eal
where we have used the identity 
$\vartheta_{2}(0)\vartheta_{4}(0)=\sqrt{kk^{\prime}}\vartheta_{3}^{2}(0)$.
We note that all the components of the Lax operator have polynomial dependence on the 
spectral parameter and with the orders of the polynomials matching the results of \sect{all_g}.

Recall that, due to the tracelessness of $L$, we have that 
\beq
\det\left(y\, I-L(w,\bar{w};\lambda)\right)=0
\quad\Leftrightarrow\quad
y^{2}=-\det L(w,\bar{w};\lambda)\,,
\eeq
which is the equation of the algebraic curve. It is straightforward to read off the algebraic curve from \eqn{g_1_fin}
\beq
y^{2}=\lambda^{3}+\lambda^{2}\left\{-\frac{1}{kk^{\prime}}\left(\frac{\vartheta_{2}(z)\vartheta_{4}(z)}{\vartheta_{3}(z)\vartheta_{1}(z)}\right)^{2}+\left(\frac{\vartheta_{3}(z)}{\vartheta_{1}(z)}\right)^{2}-\left(\frac{\vartheta_{1}(z)}{\vartheta_{3}(z)}\right)^{2}\right\}-\lambda\,.
\eeq
We may express the elliptic theta functions of the second and fourth kinds here in terms of the first and third kinds via the identity
\beq
\vartheta_{2}^{2}(z)\vartheta_{4}^{2}(z)=kk^{\prime}\left(\vartheta_{3}^{4}(z)-\vartheta_{1}^{4}(z)+\vartheta_{3}^{2}(z)\vartheta_{1}^{2}(z)\left[\frac{\vartheta_{2}^{2}(0)}{\vartheta_{4}^{2}(0)}-\frac{\vartheta_{4}^{2}(0)}{\vartheta_{2}^{2}(0)}\right]\right),
\eeq
which is derived from the addition theorem. From this we find the nice expression for the curve
\beq
y^{2}=\lambda\left(\lambda^{2}+\lambda\left[\frac{\vartheta_{4}^{2}(0)}{\vartheta_{2}^{2}(0)}-\frac{\vartheta_{2}^{2}(0)}{\vartheta_{4}^{2}(0)}\right]-1\right).
\eeq
A final identity giving the position of the cuts of the Riemann surface associated to the theta functions
\beq
\left(\frac{\vartheta_{2}(0)}{\vartheta_{4}(0)}\right)^{2}=\lambda_1\,,
\eeq
leads to
\beq\label{g_1_c}
y^{2}=\lambda\left(\lambda-\lambda_1\right)\left(\lambda+1/\lambda_{1}\right).
\eeq

We have thus recovered the genus-one algebraic curve explicitly from the Lax operator. 
We recover this curve via an alternative approach in \app{fact}.
In \cite{janik,dekel,klose} several algebraic curves of genus-zero or genus-one 
in $\bH_3$ were constructed. 
It is straightforward to see that the curve \eqn{g_1_c} agrees with the curves found in 
these references.

\section{Discussion}

We have constructed a Lax operator and the algebraic curve for the most general 
minimal surface solution of \cite{ishizeki}, which holographically describe Wilson loop 
operators in $\bR^2$. This was done via a 
modification of the prescription of \cite{janik}. We found that the curve is given by the 
the same hyperelliptic curve defining the Riemann surface.

Though there are ambiguities in defining the Lax operator and the fact that it is local, 
it is still natural to expect the resulting curve to be unique, and not dependent on these 
ambiguities. This suggest that this curve plays the same role as that normally derived 
from the monodromy matrix.

It would of course be interesting to rederive these results from a monodromy matrix 
on the open string surface (including appropriate reflections from the boundaries). 
Monodromies have been 
constructed for open strings with certain integrable D-brane boundary conditions \cite{bajnok}, 
but no such construction has been successfully applied to strings ending on the 
boundary of $AdS$.

An important difference between the closed and open string cases is that given an 
algebraic curve there are many minimal surfaces associated to it, not related to each-other 
by a global symmetry. This is evident in the construction of \cite{ishizeki}, where the solution 
is given in terms of a curve, a continues spectral parameter on the unit circle and a discrete 
choice among different possible boundaries for the string. We have seen it 
from the opposite point of view, where we indeed found the same algebraic curve for 
all these surfaces, irrespective of the value of the spectral parameter, which is just 
the phase (or imaginary part) of our full complex spectral parameter.

\section*{Acknowledgements}
We would like to thank Martin Kruczenski for invaluable discussions and sharing his 
computer code with us. 
N.D. is grateful for the hospitality of APCTP, of Nordita and CERN (via the 
CERN-Korea Theory Collaboration) during the course of this work. 
The research of N.D. is underwritten by an STFC advanced fellowship. The 
CERN visit was funded by the National Research Foundation (Korea).

\appendix

\section{Taylor expansion of theta functions}
\label{app:tay}

For our analysis in \sect{all_g} we need to understand the behaviour of theta 
functions of the form  $\theta(z\pm\int_{0}^{\lambda})$ with $\lambda$ near a branch 
point $\lambda_i$. 
As can be seen from \eqn{explicit-omega} (see also \cite{kski}), the differentials are 
best expressed in terms of the variables $y=-i\sqrt{\lambda-\lambda_{i}}$, and then 
have a finite limit at $\lambda=\lambda_i$. It is straightforward to see that the integral 
$\int_{0}^{\lambda}$ is an odd function of $y$. The Taylor expansion about $y=0$ gives
\beq
\textstyle
\theta(z\pm\int_{0}^{\lambda})=\theta(z\pm\int_{0}^{\lambda_{i}})
\pm y\,D_{\lambda_{i}}\theta(z\pm\int_{0}^{\lambda_{i}})+\mathcal{O}(y^{2})\,.
\eeq

We are particularly interested in the following expansion about $\lambda=0$
\beq
\textstyle
\hat{\theta}(z+\int_{0}^{\lambda})\theta(z-\int_{0}^{\lambda})
-\theta(z+\int_{0}^{\lambda})\hat{\theta}(z-\int_{0}^{\lambda})
=2y\big(\theta(z)D_{0}\hat{\theta}(z)-\hat{\theta}(z)D_{0}\theta(z)\big)+\mathcal{O}(y^{3})\,,
\eeq
which we see behaves as $\sqrt{\lambda}$.

Similarly, for large $\lambda$ we may expand about $w=-i/\sqrt{\lambda}$ 
and it is straightforward to see that
\beq
\textstyle
\hat{\theta}(z+\int_{0}^{\lambda})\theta(z-\int_{0}^{\lambda})
-\theta(z+\int_{0}^{\lambda})\hat{\theta}(z-\int_{0}^{\lambda})
=2w\big(\theta(z)D_{\infty}\hat{\theta}(z)-\hat{\theta}(z)D_{\infty}\theta(z)\big)+\mathcal{O}(w^{3})\,,
\eeq

\section{The zeros of $\hat{\theta}(2\int_{0}^{\lambda})$}
\label{app:zeros}

We now show that $\hat{\theta}(2\int_{0}^{\lambda})$ has triple zeros at the 
\emph{odd finite branch points}, $\{\lambda_i^-\}$. 
As in the previous appendix, we may Taylor expand $\hat{\theta}(2\int_{0}^{\lambda})$ about a 
branch point $\lambda_i$ ({\em i.e.}, $y=0$)
\beq
\textstyle
\hat{\theta}(2\int_{0}^{\lambda})
=2y\,D_{\lambda_i}\hat{\theta}(2\int_{0}^{\lambda_{i}})
+\mathcal{O}(y^{3})\,,
\eeq
This is a simple zero if $D_{\lambda^-_i}\hat{\theta}(2\int_{0}^{\lambda^-_{i}})$ is finite and 
at least a triple zero (since the function is odd) if it vanishes. 
The path $2\int_0^{\lambda_i}$ defines a closed cycle and we may write 
$2\int_0^{\lambda_i}\equiv m+\Pi n$ for $m,n\in\mathbb{Z}$.
Using the theta function identities from the main text and recalling that 
$2\int_0^\infty=l+\Pi k$, 
one may write
\beq
\hat{\theta}(z+2{\textstyle\int_0^{\lambda_i}})
=C\exp\left\{\pi i\left(k^{\intercal}-2n^{\intercal}\right)z\right\}\theta(z+{\textstyle\int_0^\infty})\,,
\eeq
where $C$ is a nonzero constant. 
Since $\theta(\int_0^\infty)$ vanishes,
It is thus apparent that
\beq
D_{\lambda_i}\hat{\theta}(2{\textstyle\int_0^{\lambda_i}})
=D_{\lambda_i}\hat{\theta}(z+2{\textstyle\int_0^{\lambda_i}})\Big|_{z=0}
=C\exp\left\{\pi i\left(l^{\intercal}-2n^{\intercal}\right)z\right\}
D_{\lambda_i}\theta(z+{\textstyle\int_0^\infty})\Big|_{z=0}
=CD_{\lambda_i}\theta({\textstyle\int_0^\infty})\,.
\eeq
In particular, $D_{\lambda_i}\theta(\int_0^\infty)=0$ is equivalent to $D_{\lambda_i}\hat{\theta}(2\int_0^{\lambda_i})=0$.
It therefore suffices to show that the directional derivative, 
$D_{\lambda^{-}_{i}}\theta(\int_0^\infty)$, vanishes. To do so, we shall employ 
\emph{Fay's trisecant identity}, 
see {\em e.g.}, \cite{mfordii}. A corollary of the trisecant identity states that
\beq\label{tris_1}
D_{\lambda}\ln\frac{\theta(z)}{\theta(z+\int_{\eta}^{\rho})}
=-D_{\lambda}\ln\frac{\theta(a+\int_{\eta}^{\lambda})}{\theta(a+\int_{\rho}^{\lambda})}
-\frac{D_{\lambda}\theta(a)\theta(a+\int_{\rho}^{\eta})}{\theta(a+\int_{\rho}^{\lambda})
\theta(a+\int_{\lambda}^{\eta})}
\frac{\theta(z+\int_{\eta}^{\lambda})\theta(z+\int_{\lambda}^{\rho})}{\theta(z)
\theta(z+\int_{\eta}^{\rho})}\,,
\eeq
where $a$ is a non-singular zero of theta. Let us choose $\rho=0$, $a=\int_{0}^\infty$ and $z=\int^\infty_0+\int^{\lambda}_0$:
\bal\label{tris_con}
D_{\lambda}\ln\frac{\theta(\int^\infty_0+\int^{\lambda}_0)}{\theta(\int^\infty_0+\int^{\lambda}_\eta)}
=-D_{\lambda}\ln\frac{\theta(\int^\infty_0+\int^{\lambda}_\eta)}{\theta(\int^\infty_0+\int^{\lambda}_0)}
-\frac{D_{\lambda}\theta(\int_0^{\infty})\theta(\int_{0}^\infty+\int_0^\eta)}{\theta(\int_0^\infty+\int_0^{\lambda})\theta(\int_{0}^\infty+\int_{\lambda}^\eta)}
\frac{\theta(\int^\infty_\eta+2\int^{\lambda}_0)\theta(\int_0^\infty)}{\theta(\int^\infty_0+\int_0^{\lambda})\theta(\int^\infty_0+\int^{\lambda}_\eta)}\,.
\eal
It is apparent that the left-hand side cancels with the first term of the right-hand side. 
Let us examine the analytic structure of the second term on the right-hand side. 
Taking $\lambda$ and $\eta$ to be regular points on the surface we have a simple 
zero due to the $\theta(\int_0^\infty)$ factor in the numerator. The identity is thus satisfied. 

The situation is different when $\lambda$ is an odd finite branch point, since 
$\theta(\int_0^\infty+\int_0^{\lambda_i})$ in the denominator vanishes.
This may be seen by noting that $\theta(z)=0$ if and only if $z\in W_{g-1}+\kappa$, 
where $W_{g-1}$ is the image of integral divisors of degree $g-1$ on $\Sigma$ under the 
Abel-Jacobi mapping and $\kappa$ is the \emph{vector of Riemann constants} \cite{farkas}. 
The vector of Riemann constants may be given by the image of the divisor of the odd branch 
point characteristics under the Abel-Jacobi mapping, {\em i.e.}, 
$\kappa=\int_{0}^{\infty}+\sum_{j}^{g-1}{\textstyle\int_{0}^{\lambda_{j}^-}}$. We may write
\beq
{\textstyle\int^{\lambda_i}_0+\int^\infty_0}
={\textstyle\int^{\lambda_i}_0}
-\sum_{j}^{g-1}{\textstyle\int^{\lambda^{-}_{j}}_{0}}
+\kappa\,.
\eeq
It is apparent that ${\textstyle\int^{\lambda_i}_0}
-\sum_{j}^{g-1}{\textstyle\int^{\lambda^{-}_{j}}_{0}}\in W_{g-1}$ if and only if 
$\lambda_i$ is an odd \emph{finite} branch point. 

Thus, in order for the identity \eqn{tris_con} to be satisfied for $\lambda=\lambda_i^-$, we find
\beq
D_{\lambda_i^-}\theta({\textstyle\int^\infty_0})
=0\,,
\eeq
as required, and indeed it is at least a double zero.
We emphasize that the 
same procedure does not apply for $D_{\infty}\theta(\int_0^\infty)$, or indeed for the even branch points.

We therefore find that $\hat{\theta}(2\int_{0}^{\lambda})$ has a zero of order (at least) three 
at the odd finite branch points. As there are $g-1$ finite odd 
branch points and $g+1$ other branch points which are also zeros, we find a total of 
at least $3(g-1)+(g+3)=4g$, zeros. But as stated at the end of \sect{theta}, this is exactly 
the number of zeros of $\hat{\theta}(2\int_{0}^{\lambda})$, so we have identified the 
correct order of them all.

\section{Equivalence of the sigma-model and Pohlmeyer reduced Lax operators}
\label{app:lax}

It is straightforward to see that the currents \eqn{gen_mc} are given in terms of the 
reduced currents \eqn{param} via
\beq
\mathcal{J}(x)=\frac{1}{1-x}(h^{\dagger})^{-1}(j+\bar{j}^{\dagger})h^{\dagger}\,,\qquad\bar{\mathcal{J}}(x)=\frac{1}{1+x}(h^{\dagger})^{-1}(\bar{j}+j^{\dagger})h^{\dagger}\,.
\eeq
with $h$ from \eqn{decomp}. 
Taking the Lax connection $M$ in \eqn{lax_eqn} to be given by the generalized 
Maurer-Cartan forms \eqn{gen_mc} the holomorphic part of the Lax equation may be written as
\beq\label{lax_i}
\partial L=\frac{1}{1-x}[L,(h^{\dagger})^{-1}(j+\bar{j}^{\dagger})h^{\dagger}]\,.
\eeq
Let us define $\cL\equiv h^{\dagger}L(h^{\dagger})^{-1}$. Then \eqn{lax_i} is equivalent to
\beq
\partial\cL=\frac{1}{1-x}\left([\cL,j]+x[\cL,\bar{j}^{\dagger}]\right),
\eeq
which is none other than
\beq
\partial\cL=[\cL,J].
\eeq
Therefore, if we solve for $\cL$ we may obtain $L$ by simple gauge transformation 
and furthermore the algebraic curve of $L$ and $\cL$ are identical.

\section{Factorized string method for genus-one}
\label{app:fact}

We have constructed the algebraic curve for all-genus. This was done for a Lax operator 
with respect to the Pohlmeyer connection. We now construct the Lax operator with 
respect to the sigma-model connection, for the case of genus-one. The resulting expression 
will be given in terms of the sigma-model spectral parameter $x$, rather than the 
Pohlmeyer spectral parameter $(1+x)/(1-x)$.

The factorized string method of \cite{dekel} separates the functional dependence of the 
connections on the two world-sheet coordinates $\sigma$ and $\tau$ (where 
$w=\sigma+i\tau$). This can be implemented for the genus-one solutions where as we 
will show, the Maurer-Cartan forms may be expressed as
\beq\label{fact}
\mathcal{J}_{\sigma}(\sigma,\tau)
=S^{-1}(\sigma)\mathcal{J}_{\sigma}^{0}(\tau)S(\sigma)\,,
\qquad
\mathcal{J}_{\tau}(\sigma,\tau)
=S^{-1}(\sigma)\mathcal{J}_{\tau}^{0}(\tau)S(\sigma)\,.
\eeq
In the case of interest $S\in SL(2;\mathbb{C})/SU(2)$. As a consequence of this 
factorization, a Lax operator may be defined as 
$L\equiv\mathcal{J}_{\sigma}(x)-\mathcal{J}_{S}^{L}$. Here 
$\mathcal{J}_{\sigma}(x)
\equiv\frac{1}{1-x^{2}}(\mathcal{J}_{\sigma}-ix\mathcal{J}_{\tau})$ 
and $\mathcal{J}_{S}^{L}\equiv S^{-1}\partial_{\tau}S=\text{constant}$.%
\footnote{This is consistent with the definition of the generalized currents, 
\eqn{gen_mc}, in terms of $w$ and $\bar w$.} 
Note that this Lax operator is with respect to the generalized Maurer-Cartan 
connection, {\em c.f.}, the discussion in \sect{all_g}, {\em i.e.}, it satisfies
\beq
\partial_{\sigma}L=[L,\mathcal{J}_{\sigma}(x)]\,,\qquad\partial_{\tau}L=[L,\mathcal{J}_{\tau}(x)]\,.
\eeq

The first step is to write down the explicit expressions for the Maurer-Cartan forms 
for these solutions. It follows from the definitions of the currents and \eqn{h} that
\bal
\mathcal{J}_{\sigma}^{\star}
&=\frac{\theta^{2}(\int_{0}^{\lambda})}{D_{\infty}\hat{\theta}(0)\,\hat{\theta}(2\int_{0}^{\lambda})}
\frac{1}{\hat{\theta}^{2}(z)}
\begin{pmatrix}
\lambda\,\theta_{+}\theta_{-}-\hat{\theta}_{+}\hat{\theta}_{-} 
& \left(\lambda\,\theta^{2}_{-}+\hat{\theta}^{2}_{-}\right)e^{-2\tilde{\mu}\sigma-2\tilde{\nu}\tau}
\\\left(-\lambda\,\theta^{2}_{+}-\hat{\theta}^{2}_{+}\right)e^{2\tilde{\mu}\sigma+2\tilde{\nu}\tau}
& -\lambda\,\theta_{+}\theta_{-}+\hat{\theta}_{+}\hat{\theta}_{-}
\end{pmatrix},
\\
\mathcal{J}_{\tau}^{\star}
&=\frac{\theta^{2}(\int_{0}^{\lambda})}{D_{\infty}\hat{\theta}(0)\,\hat{\theta}(2\int_{0}^{\lambda})}
\frac{i}{\hat{\theta}^{2}(\zeta)}
\begin{pmatrix}
-\lambda\,\theta_{+}\theta_{-}-\hat{\theta}_{+}\hat{\theta}_{-} 
& \left(-\lambda\,\theta^{2}_{-}+\hat{\theta}^{2}_{-}\right)e^{-2\tilde{\mu}\sigma-2\tilde{\nu}\tau}
\\\left(\lambda\,\theta^{2}_{+}-\hat{\theta}^{2}_{+}\right)e^{2\tilde{\mu}\sigma+2\tilde{\nu}\tau} 
& \lambda\,\theta_{+}\theta_{-}+\hat{\theta}_{+}\hat{\theta}_{-}
\end{pmatrix},
\eal
where we wrote the complex conjugates of the currents for presentational purposes and we defined 
$\tilde{\mu}(\lambda)\sigma+\tilde{\nu}(\lambda)\tau\equiv\mu(\lambda)z+\nu(\lambda)\bar{z}$,%
\footnote{Again, the specific forms of $\tilde{\mu}$ and $\tilde{\mu}$ are not important,
 only that they are functions of the spectral parameter.} 
and we have introduced the shorthand, $\theta_{\pm}\equiv\theta(z\pm\int_{0}^{\lambda})$ 
and $\hat{\theta}_{\pm}\equiv\hat{\theta}(z\pm\int_{0}^{\lambda})$. 
We treat $\lambda$ as a constant here.

We have thus far failed to mention anything about the form of $z(w,\bar{w})$. 
It is related to the world-sheet coordinates by \cite{ishizeki}
\beq\label{}
z\equiv2\left[\omega(\infty)w+\omega(0)\bar{w}\right]\,,
\eeq
which for the genus-one case, reduces to
\beq
z=4i\omega(\infty)\tau\,.
\eeq
In particular, this ensures that none of the theta functions are a function of $\sigma$. This may be exploited to factorize the currents with respect to $\sigma$. Consider the matrix
\beq
S(\sigma)=
\begin{pmatrix}
e^{\tilde{\mu}\sigma} & 0\\
0 & e^{-\tilde{\mu}\sigma}
\end{pmatrix}
\in SL(2;\mathbb{C})/SU(2)\,.
\eeq
It is apparent that this matrix satisfies \eqn{fact}, with
\beq
\mathcal{J}_{\sigma}^{0\star}=\frac{\theta^{2}(\int_{0}^{\lambda})}{D_{\infty}\hat{\theta}(0)\,\hat{\theta}(2\int_{0}^{\lambda})}\frac{1}{\hat{\theta}^{2}(z)}
\begin{pmatrix}
\lambda\,\theta_{+}\theta_{-}-\hat{\theta}_{+}\hat{\theta}_{-} & \left(\lambda\,\theta^{2}_{-}+\hat{\theta}^{2}_{-}\right)e^{-2\tilde{\nu}\tau}\\
\left(-\lambda\,\theta^{2}_{+}-\hat{\theta}^{2}_{+}\right)e^{2\tilde{\nu}\tau} & -\lambda\,\theta_{+}\theta_{-}+\hat{\theta}_{+}\hat{\theta}_{-}
\end{pmatrix}
\,,
\eeq
\beq
\mathcal{J}_{\tau}^{0\star}=\frac{\theta^{2}(\int_{0}^{\lambda})}{D_{\infty}\hat{\theta}(0)\,\hat{\theta}(2\int_{0}^{\lambda})}\frac{i}{\hat{\theta}^{2}(\zeta)}
\begin{pmatrix}
-\lambda\,\theta_{+}\theta_{-}-\hat{\theta}_{+}\hat{\theta}_{-} & \left(-\lambda\,\theta^{2}_{-}+\hat{\theta}^{2}_{-}\right)e^{-2\tilde{\nu}\tau}\\
\left(\lambda\,\theta^{2}_{+}-\hat{\theta}^{2}_{+}\right)e^{2\tilde{\nu}\tau} & \lambda\,\theta_{+}\theta_{-}+\hat{\theta}_{+}\hat{\theta}_{-}
\end{pmatrix}
\,.
\eeq
Additionally
\beq
\mathcal{J}_{S}^{L}=\tilde{\mu}
\begin{pmatrix}
1 & 0\\
0 & -1
\end{pmatrix}
=\text{constant}\,.
\eeq

As $L$ is traceless, the spectral curve may be given by
\beq
y^{2}=\frac{1}{2}\tr L^{2},
\eeq
or more explicitly
\beq
\begin{aligned}
y^{2}=&\,\frac{1}{2(1-x^{2})^{2}}\tr\Big[(\mathcal{J}_{\sigma}-\mathcal{J}_{S}^{L})^{2}+x\left\{2i(\mathcal{J}_{S}^{L}-\mathcal{J}_{\sigma})\mathcal{J}_{\tau}\right\}+x^{2}\left\{2(\mathcal{J}_{\sigma}-\mathcal{J}_{S}^{L})\mathcal{J}_{S}^{L}-\mathcal{J}_{\tau}^{2}\right\}\\
&{}+x^{3}\left\{-2i\mathcal{J}_{S}^{L}\mathcal{J}_{\tau}\right\}
+x^{4}\left\{(\mathcal{J}_{S}^{L})^{2}\right\}\Big].
\end{aligned}
\eeq
The calculation of the coefficients is straightforward following \sect{g_1}. The result is
\beq
\begin{aligned}
y^{2}=\frac{1}{(1-x^{2})^{2}}&\left\{2\left(\lambda-\frac{1}{\lambda}\right)-2\left(a-\frac{1}{a}\right)+x\left[-4\left(\lambda-\frac{1}{\lambda}\right)\right]+x^{2}\left[4\left(a-\frac{1}{a}\right)\right]\right.\\
&\left.+x^{3}\left[4\left(\lambda-\frac{1}{\lambda}\right)\right]+x^{4}\left[-2\left(\lambda-\frac{1}{\lambda}\right)-2\left(a-\frac{1}{a}\right)\right]\right\}
\end{aligned}\,,
\eeq
which is birationally equivalent to
\beq
y^{2}=\lambda\frac{1+x}{1-x}-\frac{1}{\lambda}\frac{1-x}{1+x}-a+\frac{1}{a}\,.
\eeq

We see that solutions with different values of $\lambda$ lead to birationally equivalent 
curves. Let us absorb $\tilde\lambda\equiv\lambda\frac{1+x}{1-x}$ and 
$\tilde{y}\equiv y/\tilde\lambda$ which brings the curve to the standard form
\beq
\tilde y^{2}=\tilde\lambda\left(\tilde\lambda-a\right)\left(\tilde\lambda+\frac{1}{a}\right).
\eeq

\bibliography{ref_nd}
\end{document}